\documentclass[12pt]{iopart}

\usepackage[]{units}
\usepackage{acronym}
\usepackage[T1]{fontenc}
\usepackage{lettrine} 
\usepackage{gensymb} 
\usepackage{float} 
\usepackage{nomencl}
\usepackage{caption}
\usepackage{subcaption}
\usepackage{ragged2e}
\usepackage{multirow}
\newcommand{\ssymbol}[1]{^{\@fnsymbol{#1}}}
\makeatother
\usepackage{graphicx}
\usepackage[colorinlistoftodos]{todonotes}
\usepackage{hyperref} 
\usepackage{booktabs,caption}
\usepackage[flushleft]{threeparttable}
\hypersetup{colorlinks = true,linkcolor = blue,anchorcolor =red,citecolor = blue,filecolor = red,urlcolor = blue}
\graphicspath{ {./img/} }

\acrodef{IR}[IR]{Infra-Red}
\acrodef{INFN}[INFN]{Istituto Nazionale di Fisica Nucleare}
\acrodef{PCTO}[PCTO]{Percorsi per le Competenze Trasversali e per l'Orientamento}
\acrodef{PLS}[PLS]{Piano Lauree Scientifiche}
\begin{document}

\title[Teaching Physics by Arduino during COVID-19 Pandemic: Oscillation of a simple pendulum]{Teaching Physics by Arduino during COVID-19 Pandemic: Oscillation of a simple pendulum}

\author{Fausto Casaburo}

\address{Sapienza Università di Roma, Dipartimento di Fisica}
\address{Istituto Nazionale di Fisica Nucleare (INFN) Sezione Roma}
\ead{fausto.casaburo@uniroma1.it-fausto.casaburo@roma1.infn.it}
\vspace{10pt}
\begin{indented}
\item[]August 2021
\end{indented}

\begin{abstract}
The COVID-19 impacted on teaching worldwide; indeed, both schools and universities had to shift from face-to-face to distance teaching organizing on-line lectures. Thanks to easily accessible materials, smartphones physics apps, on-line tools and devices, it's possible to perform laboratory practice even in this period. In this paper, a method to measure the gravitational acceleration by oscillation of a simple pendulum, using Arduino board, is presented.

\end{abstract}

\vspace{2pc}

%
%
%


\tableofcontents

\section*{Introduction}\label{Introduction}
\addcontentsline{toc}{section}{Introduction}
\lettrine[nindent=0em,lines=3]{T}he COVID-19 pandemic impacted on teaching worldwide. Despite that, to overcame the problem, the educational
system replaced face-to-face by distance learning \cite{PhysRevPhysEducRes.17.010117, casaburo2021teaching, casaburo2021teaching2}. 

In particular, to preserve laboratory courses, during the COVID-19 pandemic, many schools and universities proposed to perform scientific experiments at home \cite{doi:10.1119/5.0020515}.

For example, in Italy, the Lab2Go project \cite{Lab2Gowebpage,Lab2GoWiki,ORGANTINI2017PRO, andreotti2021il, astone2021studio} organized on-line  seminars showing physics experiments that can be made at home using easily accessible materials and exploiting resources as the Arduino board \cite{Arduino}.

Arduino is an open source platform made of electronic boards, sensors and expansion boards. Thanks to its versatility and low cost, 
Arduino is an ideal base on which to construct datalogging sensors or control devices that can be used to perform phyics experiments \cite{Kinchin_2018}. Its usage also allows to acquire additional competences, as for example coding and programming \cite{Organtini_2018, Organtini_fisica_arduino}. The original Arduino board can be bought for as low as about few tens of euro; moreover, there are many cheapest clones. Both the board and the sensors can be easily bought on-line or in electronics shops \cite{Kinchin_2018, Organtini_fisica_arduino}. There are also many kits including the board and most common sensors available for just 50-60 euro. Thanks to the low cost, the Arduino board and related components can be bought directly by students or by schools/universities to be provided to students. Moreover, even if you are a beginner, there are many introductory free on-line tutorials which give a basic though thorough introduction to the Arduino boards \cite{Kinchin_2018}.

In this paper an Arduino-based physics experiment regarding of oscillation of a simple pendulum will be presented. It consists in measuring the oscillation period by Arduino to estimate the gravitational acceleration value.

Even if this is a very common experiment, it is usually realized by expansive photogate sensors in physics laboratories \cite{Yulkifli_2018}. In this paper it will be shown how to replace the photogate by cheaper \ac{IR} pair sensors and a hand-made U-shape support. 
 
 The experiment can be proposed both to high school and university students.

\section{Theory}\label{Theory}
\begin{figure}[H]
\centering
\includegraphics[width=3in]{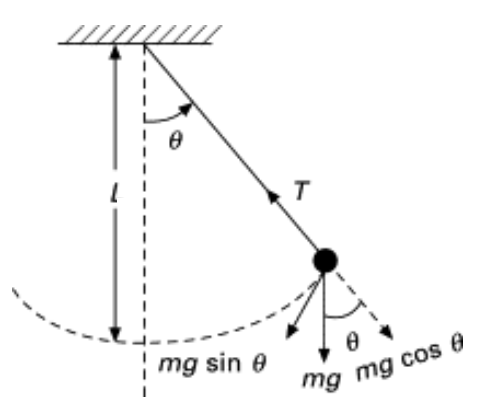}\DeclareGraphicsExtensions.
\caption{Schematization of a simple pendulum \cite{RAJASEKARAN20099}.}
\label{fig:simple_pendulum}
\end{figure}

The simple pendulum (Fig. \ref{fig:simple_pendulum}) is a mechanical system of length $l$, made of a point-mass $m$ suspended by means of light inextensible string from a fixed support.  

The equilibrium position is when the string hangs vertically. When displaced to an initial angle $\theta$ and released, the pendulum will swing back and forth due to gravitational acceleration. If there aren't other forces acting on the pendulum, the motion is periodic of period $T$. 

Moreover, for small $\theta$, the period is given by:

\begin{equation}
T=2\pi\sqrt{\frac{l}{g}}
\label{eq:T_pendulum}
\end{equation}

where $g$ is the gravitational acceleration (the average value on the Earth is $g=\unitfrac[9.80665]{m}{s^{2}}$ \cite{Valore_g_doi:10.1063/1.555817}). 

By equation \ref{eq:T_pendulum}, it follows that measuring the period $T$ and the length $l$ of the pendulum it's possible to estimate the gravitational acceleration.

\section{Experimental setup}\label{Setup}
\begin{figure}[H]
\centering
\includegraphics[width=3in]{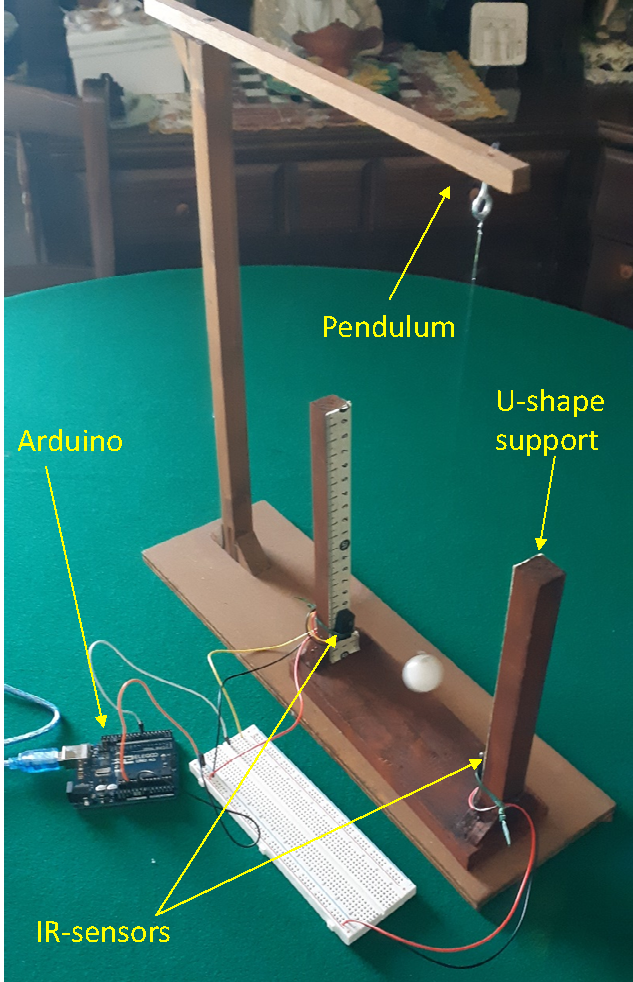}\DeclareGraphicsExtensions.
\caption{Experimental setup. }
\label{fig:exp_setup}
\end{figure}

The experimental setup is shown in  Fig. \ref{fig:exp_setup} and it is made of mechanic and electronic systems. The first one consist of a hand-made pendulum made of a wood support, a nylon string and a spherical mass. The second one consist of an Arduino UNO R3 board, an \ac{IR} trasmitter-receiver pair sensors, a breadboard, Dupont cables, USB cable and one computer. 

The \ac{IR}  sensors are connected to a digital pin of Arduino by the Dupont cables and are supported by a hand-made U-shape support making up the hand-made photogate. This latter, in turn, is positioned close to the pendulum, so that the mass of the pendulum can moves through the \ac{IR} sensors allowing the the oscillation period measurement, as shown in Fig. \ref{fig:exp_setup}. 

The Arduino board is connected to the computer for data acquisition by the USB cable.

\section{Experimental procedure}\label{Procedure}

\begin{figure}[H]
\centering
\includegraphics[width=3in]{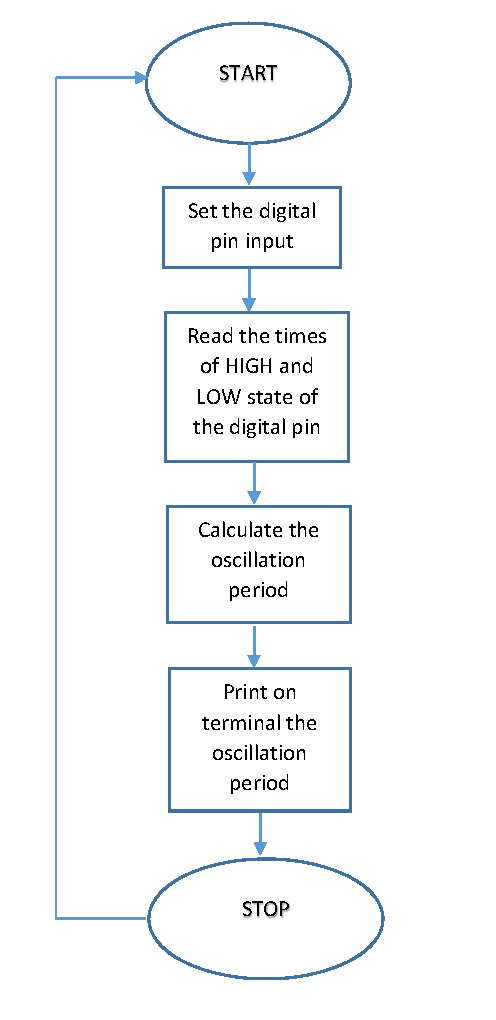}\DeclareGraphicsExtensions.
\caption{Programming flow chart for measuring the period of the pendulum by Arduino.}
\label{fig:diagramma}
\end{figure}

The sketch for Arduino allows the user to measure the oscillation period (Fig. \ref{fig:diagramma}). 

The \ac{IR} light is constantly emitted from the transmitter of the \ac{IR} pair sensors and, if there aren't obstacles, it will hit the \ac{IR} receiver and it is read by the Arduino board as a HIGH state of the pin whose the sensor is connected. When the pendulum's mass  moves through the beam light, this one is interrupted, therefore the pin state of the sensor will be switched to LOW state.  
The period is measured as the time between two consecutive passages of the pendulum's mass through the \ac{IR} sensors and it's printed on the terminal. 

The length $l$ of the pendulum is the distance from the suspension point to the barycenter of the mass $m$ (a sphere). Therefore, it has been measured the diameter $d=\unit[\left(20.00\pm0.05\right)]{mm}$ of the sphere by a calipers and the length $l^{*}$ of the string by a meterstick (sensitivity $\pm\unit[1]{mm}$ ). 
The total length $l$ of the pendulum is 
\begin{equation}
l=l^{*}+\frac{d}{2}
\label{eq:linear_function}
\end{equation}
The periods for several pendulum lengths have been measured and analyzed. For each pendulum length it has been measured the periods $T_n$ of the first 20 oscillations, then it has been calculated the average period $\left\langle T\right\rangle$ and its uncertainty assuming a Gaussian distribution. 

Data of $\left\langle T\right\rangle ^{2}$ in fuction of $l$ have been interpulated, using ROOT \cite{ROOT}, by the linear function:

\begin{equation}
y=kl
\label{eq:linear_function}
\end{equation}

where $y=\left\langle T\right\rangle ^{2}$ and $k=\frac{4\pi^{2}}{g}$. Therefore, the gravitational acceleration is given by:

\begin{equation}
g=\frac{4\pi^{2}}{k}
\label{eq:g_slope}
\end{equation}

Uncertainties of $\left\langle T\right\rangle ^{2}$ and $l$ have been calculated by the well-known rules of uncertainties propagation.

\section{Results}\label{Results}

\begin{figure}[H]
\centering
\includegraphics[width=5in]{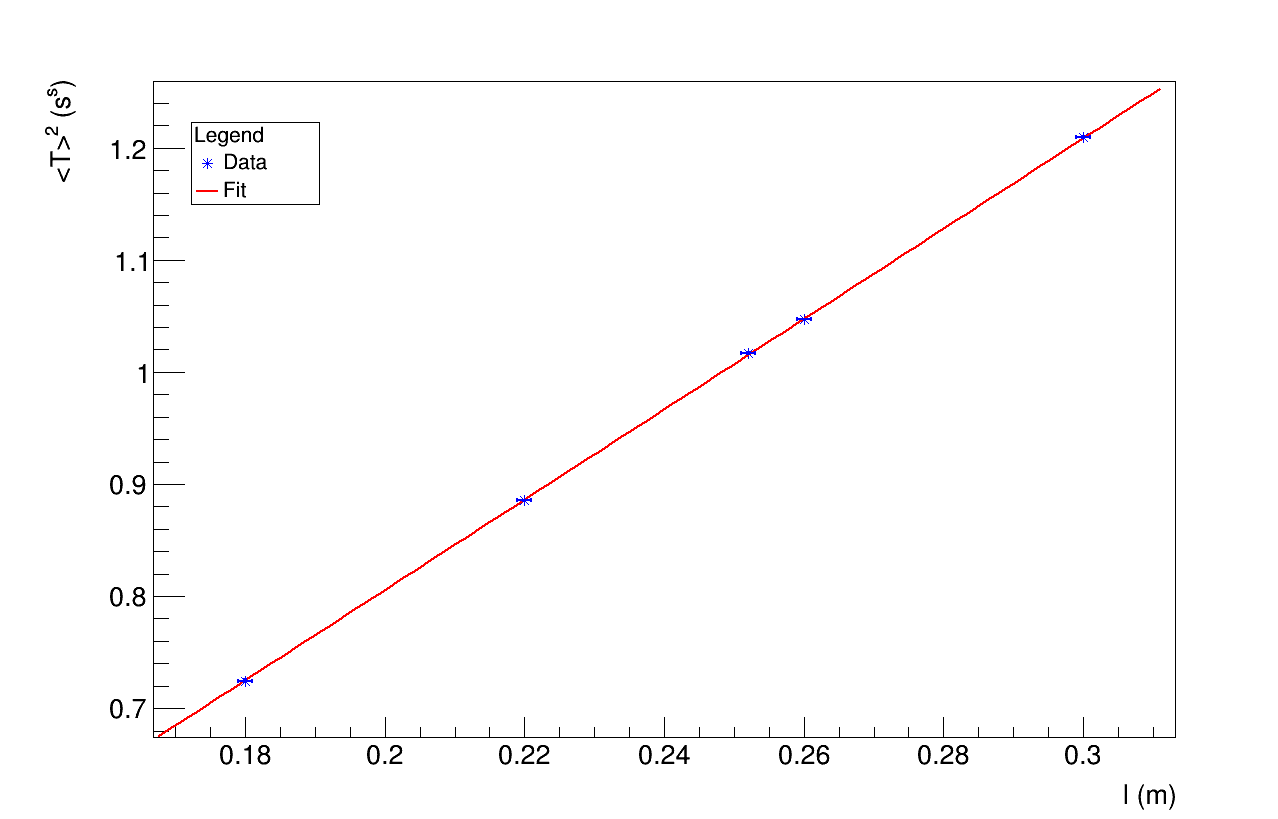}\DeclareGraphicsExtensions.
\caption{Squared oscillation period $\left\langle T\right\rangle ^{2}$ in function of the pendulum length $l$.}
\label{fig:fit}
\end{figure}

The graph of $\left\langle T\right\rangle ^{2}$ in function of  $l$ is shown in Fig. \ref{fig:fit}. 

The linear interpulation results in a slope $k$ given in Tab. \ref{tab:results} with the experimental gravitational acceleration value (eq. \ref{eq:g_slope}). 

\begin{table}[H]
\centering
 \caption{ Fit result and experimental $g$ value.}
 \label{tab:results} 
  \begin{threeparttable}
     \begin{tabular}{ll}
        \toprule
        \(\unit[m]{\left(\unitfrac{s^{2}}{m}\right)}\)  &  \(\unit[g_{exp}]{\left(\unitfrac{m}{s^{2}}\right)}\) \\
        \midrule
	         $4.0298\pm0.0074$   &  $9.797\pm0.018$\\
        \bottomrule
     \end{tabular}
  \end{threeparttable}
\end{table}

The obtained $g_{exp}$ value is well compatible with the local $g_{loc}=\unitfrac[9.8029]{m}{s^{2}}$ \cite{local_g} within $1\sigma$.

\section*{Conclusions}
\addcontentsline{toc}{section}{Conclusions}
Due to COVID-19 pandemic, students couldn't access to schools and universities laboratories. To overcame this problem it's useful organizing laboratory
activities made at home using easily accessible materials and exploiting resources as smartphones physics apps, on-line tools and devices, as for example Arduino.

In this paper it has been shown a technique to study the oscillation of a simple pendulum and to measure the gravitational acceleration using  Arduino and a cheap hand-made photogate. Beyond the numerical result, the article goal is to encourage teachers to propose the experiment to their students in order to carry on the laboratory practice even in this pandemic period. 

Lastly, thanks to its low-cost, the usage of Arduino for physics experiments can also be useful in school laboratories not adequately equipped (obsolete or non-functioning instrumentation, poor assortment, lack in maintenance, missing catalog) even when the COVID-19 emergency is over.

\section*{Acknowledgements}
\addcontentsline{toc}{section}{Acknowledgements}
The author acknowledges the Lab2Go- Fisica collaboration in particular, professors Pia Astone, Giulia De Bonis, Riccardo Faccini, Giovanni Organtini and Francesco Piacentini.
\section*{References}
\bibliographystyle{ieeetr}
\bibliography{mybiblio}

\acrodef{LED}[LED]{Light Emitting Diode}

\end{document}